\documentclass[journal]{IEEEtran} 
\IEEEoverridecommandlockouts
\usepackage{cite}
\usepackage{amsmath,amssymb,amsfonts,amsthm}
    \theoremstyle{definition}
    \newtheorem{example}{Example}
    \newtheorem{theorem}{Theorem}
\usepackage{graphicx}
\usepackage{textcomp}
\usepackage[dvipsnames]{xcolor}
\usepackage{balance}
\usepackage{cancel}
\usepackage{circuitikz}
\usepackage[colorlinks=true,bookmarks=false,citecolor=blue,urlcolor=blue]{hyperref}
\usepackage{soul}
\usepackage{multirow}
\usepackage{caption}
\usepackage{subfig}
\usepackage{pgfplots}
    \pgfplotsset{compat=1.17} 
    \usepgfplotslibrary{fillbetween}
\usepackage{tikz}
    \usetikzlibrary{positioning}
    \usetikzlibrary{calc}
    \usetikzlibrary{shapes}
    \usetikzlibrary{patterns} 
\newcommand{\argmax}{\mathop{\mathrm{argmax}}}
\def\Eb{E_{\text{b}}}
\def\boldv{\boldsymbol{v}}
\def\boldW{\boldsymbol{W}}
\def\boldb{\boldsymbol{b}}
\def\boldc{\boldsymbol{c}}
\def\boldr{\boldsymbol{r}}
\def\bolds{\boldsymbol{s}}
\begin{document}

\title{On the Optimality of Single-label and Multi-label Neural Network Decoders}
\author{
\IEEEauthorblockN{Yunus Can Gültekin~\IEEEmembership{Member,~IEEE}, Péter Scheepers, Yuncheng Yuan, Federico Corradi~\IEEEmembership{Member,~IEEE}, and Alex Alvarado,~\IEEEmembership{Senior Member,~IEEE}}
\thanks{This work has been partially funded by the Eindhoven Hendrik Casimir Institute (EHCI) Collaborative Projects Program.
This paper was presented in part at the International ITG Conference on Systems, Communications and Coding (SCC), Karlsruhe, Germany, March 2025~\cite{scheepers2025mlbaseddec}.}
\thanks{Y. C. Gültekin, F. Corradi, and A. Alvarado are with the Department of Electrical Engineering, Eindhoven University of Technology, Eindhoven, the Netherlands (e-mails: y.c.g.gultekin@tue.nl, f.corradi@tue.nl, a.alvarado@tue.nl).}
\thanks{P. Scheepers was with the Department of Electrical Engineering, Eindhoven University of Technology, Eindhoven, the Netherlands. He is now with the School of Electrical Engineering and Computer Science, Technical University of Berlin, Berlin, Germany (e-mail: scheepers@ccs-labs.org).}
\thanks{Y. Yuan was with the Department of Electrical Engineering, Eindhoven University of Technology, Eindhoven, the Netherlands (e-mail: y.yuan8307@outlook.com).}
}
\maketitle
\begin{abstract}
We investigate the design of two neural network (NN) architectures recently proposed as decoders for forward error correction: the so-called single-label NN (SLNN) and multi-label NN (MLNN) decoders. These decoders have been reported to achieve near-optimal codeword- and bit-wise performance, respectively. Results in the literature show near-optimality for a variety of short codes. 
In this paper, we analytically prove that certain SLNN and MLNN architectures can, in fact, always realize optimal decoding, regardless of the code.
These optimal architectures and their binary weights are shown to be defined by the codebook, i.e., no training or network optimization is required.
Our proposed architectures are in fact not NNs, but a different way of implementing the maximum likelihood decoding rule.
Optimal performance is numerically demonstrated for Hamming $(7,4)$, Polar $(16,8)$, and BCH $(31,21)$ codes.
The results show that our optimal architectures are less complex than the SLNN and MLNN architectures proposed in the literature, which in fact only achieve near-optimal performance.
Extension to longer codes is still hindered by the curse of dimensionality. 
Therefore, even though SLNN and MLNN can perform maximum likelihood decoding, such architectures cannot be used for medium and long codes.
\end{abstract}
\section{Introduction}\label{sec:intro}
\IEEEPARstart{F}{orward} error correction (FEC) is essential for reliable data transmission over noisy channels~\cite{ShannonPapers}. FEC is also a fundamental part of next-generation wireless and optical communication systems~\cite{miao2024trends,agrell2024roadmap}. 
Designing FEC codes and decoders that achieve optimum performance, with low complexity and low latency, has been the objective of coding theory for decades. %
{Maximum likelihood} decoding minimizes the error probability, and thus, it is performance-wise {optimal}~\cite{medard2020iccMLgrand}.
However, maximum likelihood decoding is impractical, as its complexity grows exponentially with increasing code blocklength $n$ for a given coding rate $R=k/n$~\cite{ml_problem}.

In practice, instead of maximum likelihood decoding, low-complexity and low-latency suboptimal decoding algorithms are preferred.
Bounded distance decoding (BDD)~\cite{berlekamp68} or decoding based on Chase-Pyndiah algorithms~\cite{chase1972class,pyndiah1998near} are widely used in combination with algebraic codes~\cite{sukmadji2022zipper} for optical communications.
Successive cancellation (SC)~\cite{Arikan2009_polarization} and SC list~\cite{tal2015list} decoding of polar codes are used in 5G cellular communications~\cite{5gnr}.
Low-density parity-check (LDPC) codes, which also find application in quantum key distribution systems~\cite{Milicevic2018} and quantum error correction~\cite{panteleev2021quantum}, are typically decoded via message-passing belief propagation~\cite{bitflip,kschischang2001factor}.

Following recent advances in machine learning, decoding FEC codes via neural network (NN) architectures has attracted considerable attention in the literature~\cite{dl-survey}.
The advantages of these data-driven decoders are especially apparent for applications where no accurate mathematical model for the channel is available, or where the corresponding traditional decoder performs poorly for the code under consideration.
As an example, consider quantum LDPC codes constructed via the Calderbank-Shor-Steane approach~\cite{calderbank1998quantum}.
The Tanner graphs of these codes have short cycles of length $4$. 
These cycles are known to significantly degrade the performance of BP and cause very high error floors~\cite{poulin2008iterative}. 
As a solution to this problem, neural BP was recently proposed in~\cite{miao2023neural}. Other examples of NN-based decoders include, e.g., \cite{liu2019neural,buchberger2020pruning,buchberger2021learned}.

Historically, NNs were first considered for decoding FEC codes in the early 1990s, e.g., in~\cite{caid1990neural}. 
NN decoders were deemed suitable only for very short codes ($n<100$) due to the curse of dimensionality~\cite{caid1990neural,wang1996artificial}.
Even considering the recent surge in NN decoders proposed in the literature, starting with~\cite{dlbasedchanneldecoding}, we see from the comprehensive survey~\cite{dl-survey} and references therein that the codes considered for NN-based decoding are almost always limited to lengths below $n=512$.

\begin{figure*}[t]
    \centering
    \includegraphics[width=0.75\textwidth]{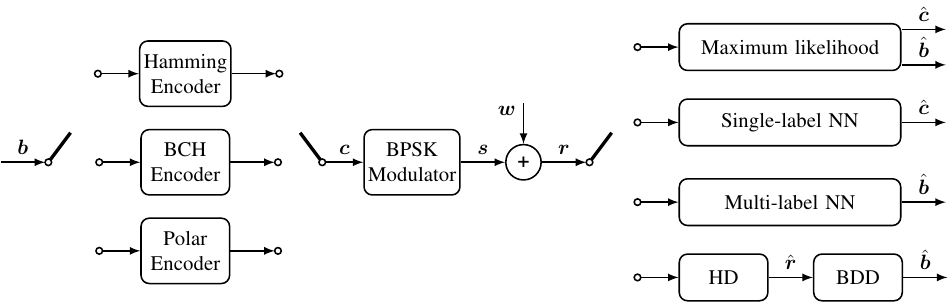}
    \caption{Communication system model considered in this work. Three possible encoders are used in combination with 4 different decoders. Maximum likelihood decoding is based on \eqref{eq:mlcodeword} or \eqref{eq:mapcri}. The SLNN and MLNN decoders are those proposed in \cite{motani2019globecom,motani2020icc,motani2020isit,motani2023trans}. BDD decoding is based on hard-decisions (HDs) $\hat{\boldsymbol{r}}$ made on $\boldsymbol{r}$.}
    \label{fig:commsyst}
\end{figure*}

Recently, single-label NN (SLNN) and multi-label NN (MLNN) decoders have been introduced~\cite{motani2019globecom,motani2020icc,motani2020isit,motani2023trans}.
These decoders have a fully-connected architecture, with a few hidden layers. SLNN and MLNN decoders have been shown in~\cite{motani2023trans} to provide codeword- and bit-wise near-optimal performance, respectively, for a variety of short ($n<34$) FEC codes. The results in~\cite{motani2023trans} were obtained by training the NNs at a fixed signal-to-noise ratio (SNR), and then, using the same NN for a wide range of SNRs. 
The NN architecture of these decoders does not rely on the knowledge of the specific code structure, i.e., they are model-free, and therefore, in principle, applicable to any code. Furthermore, these decoders are non-iterative.

The computational complexity of the SLNN and MLNN decoders becomes infeasible for large input lengths $k$.
To alleviate this problem, the iterative network pruning approach proposed in~\cite{frankle2018lottery} has been used to prune weights in the SLNN and MLNN architectures in~\cite{motani2021nips}.
With this approach, the complexity of the SLNN and MLNN decoders has been reduced by up to 97 \% without compromising the decoding performance~\cite[Sec. 6]{motani2021nips}.
The reduction in complexity has been demonstrated for the Hamming $(7, 4)$ and Polar $(16, 8)$ codes.
SLNN and MLNN with weight pruning have the potential to be scalable for longer codes, without significantly sacrificing performance~\cite{motani2021nips}.

For the aforementioned reasons, at first glance, SLNN and MLNN decoders seem to be a very interesting alternative for decoding short codes. However, only heuristic rules for the design of these decoders exist in the literature. For example, it is unclear how many hidden layers or neurons per layer these decoders should use for different codes. 
Furthermore, to the best of our knowledge, the performance of SLNN and MLNN has not been rigorously analyzed in the literature, and no comparisons to traditional decoders exist. 

In this paper, we study the design and performance of SLNN and MLNN decoders.
The first contribution of this paper is to show analytically that for certain non-fully-connected architectures with up to a single hidden layer, SLNN and MLNN decoders {\it realize} codeword- and bit-wise optimal decoding, respectively.
Moreover, the weights are binary and fully described by the codebook, i.e., no training is in fact required. 
These {\it optimal} architectures are not actually NNs but graphical representations of maximum likelihood decoding.
The second contribution is to demonstrate that these optimal non-NN architectures are significantly less complex than the non-optimal SLNN and MLNN decoders used in \cite{motani2019globecom,motani2020icc,motani2020isit,motani2023trans}. 
Finally, the third contribution is to show that already for blocklengths $n=31$, the NNs previously proposed in the literature are not competitive with respect to traditional (e.g., BDD) decoders.

This paper is organized as follows.
We provide preliminary information in Sec.~\ref{sec:preliminaries}.
SLNN and MLNN decoders are described and studied in Sec.~\ref{sec:slnnmlnn}.
Conclusions are given in Sec.~\ref{sec:conclusion}.

\section{Preliminaries}\label{sec:preliminaries}
We consider the communication system in Fig.~\ref{fig:commsyst}. The transmitter is a concatenation of an FEC encoder and a binary phase-shift keying (BPSK) mapper. 
The $k$ information bits $\boldsymbol{b} = (b_1, \dots, b_k)$ are encoded using an $(n,k)$ linear block code into a codeword $\boldsymbol{c} = (c_1, \dots, c_n)$. 
In this paper, we consider three different codes: a Hamming code, a polar code, and a Bose–Chaudhuri–Hocquenghem (BCH) code.
The codebook $\mathcal{C} = \{ \boldsymbol{c}^{(1)},\boldsymbol{c}^{(2)}, \dots, \boldsymbol{c}^{(2^k)}\}$ consists of all $2^k$ possible codewords. 
The code bits $c_i$ are mapped to BPSK symbols $s_i \in \{-1, 1\}$, via $s_i = 2c_i -1$, for $i = 1, \dots, n$.
Since the BPSK mapping is a one-to-one function, we also consider the set $\mathcal{S} = \{ \boldsymbol{s}^{(1)}, \dots, \boldsymbol{s}^{(2^k)}\}$ to be the codebook, where each element in this set is a length-$n$ vector of symbols from the set $\{-1,+1\}$.

The symbols $\boldsymbol{s} = (s_1, \dots, s_n)$ are transmitted over an additive white Gaussian noise (AWGN) channel. 
At the receiver, real-valued noisy received symbols 
\begin{align}\label{eq.awgn}
\boldsymbol{r} = \boldsymbol{s} + \boldsymbol{w},
\end{align}
are observed, where $\boldsymbol{w} = (w_1, \dots, w_n)$ is an $n$-dimensional vector of independent and identically distributed noise samples drawn from a zero-mean normal distribution with variance $\sigma^2 = {N_0}/{2}$, where $N_0$ is the power spectral density of the AWGN noise. Throughout this paper, we report results as a function of $\Eb/{N_0}$, where the energy per information bit $\Eb$ is $\Eb=1/(k/n)$, and where we assume equally likely symbols $s_i=\pm 1$.

As shown in Fig.~\ref{fig:commsyst}, the receiver estimates the information bits using an FEC decoder.
The decoder accepts as input either $\boldsymbol{r}$ (for soft-decision decoding), or hard detections $\hat{\boldsymbol{r}}$ made on $\boldsymbol{r}$ (for hard-decision decoding).
In this paper, we consider optimal soft-decision maximum likelihood decoding, the SLNN and MLNN architectures from \cite{motani2019globecom,motani2020icc,motani2020isit,motani2023trans}, and hard-decision BDD decoding. Maximum likelihood decoding and the basics required to understand SLNN and MLNN are described next.

\subsection{Maximum Likelihood Decoding}
For any $(n,k)$ linear block code whose codewords are transmitted with equal probability, the maximum likelihood decoding rule that minimizes frame (codeword) error rate (FER) estimates the transmitted codeword as 
\begin{equation}
     \hat{\boldsymbol{s}} = \argmax_{\boldsymbol{s} \in \mathcal{S}} p(\boldsymbol{r} | \boldsymbol{s}).
     \label{eq:mlcodeword}
\end{equation} 
For a memoryless channel, \eqref{eq:mlcodeword} can be rewritten as
\begin{align}
    \hat{\boldsymbol{s}} &= \argmax_{\boldsymbol{s} \in \mathcal{S}} \sum_{i=1}^{n} \log(p(r_i | s_i)).
    \label{eq:sumlogp}
\end{align}
For the AWGN channel in \eqref{eq.awgn}, the channel transition probability density function $p(r_i|s_i)$ is given by
\begin{equation}
    p(r_i|s_i) = \frac{1}{\sqrt{2 \pi \sigma^2}} \textrm{exp}\left( - \frac{(r_i - s_i)^2}{2\sigma^2}\right).
    \label{eq:awgnexp}
\end{equation}
Substituting (\ref{eq:awgnexp}) into (\ref{eq:sumlogp}) gives
\begin{align}
    \hat{\boldsymbol{s}} &= \argmax_{\boldsymbol{s} \in \mathcal{S}} \left( -\frac{n}{2} \log{(2 \pi \sigma^2 )} - \frac{1}{2\sigma^2} \sum_{i=1}^{n} (r_i - s_i)^2 \right) \label{eq:3in2}\\
    & = \argmax_{\boldsymbol{s} \in \mathcal{S}}  \sum_{i=1}^{n} 2r_i s_i - s_i^2 \label{eq:expanded}\\
    & = \argmax_{\boldsymbol{s}^{(j)} \in \mathcal{S}} 
    \sum_{i=1}^n r_ic_i^{(j)},
    \label{eq:mlcorr}
\end{align}
where \eqref{eq:expanded} follows from neglecting the terms that do not depend on~$\boldsymbol{s}$, and \eqref{eq:mlcorr} from using $s_i = 2c_i - 1$, $s_i^2 =1$, and again removing the terms that do not affect the maximization.
A decoder that performs \eqref{eq:mlcorr} is said to be codeword-wise optimal.

Similar to~\eqref{eq:mlcodeword}, the maximum likelihood decoding rule that minimizes bit error rate (BER) estimates the information bits as
\begin{equation}
    \hat{b}_{i} = \argmax_{\beta \in \{0, 1\}} p(\boldsymbol{r} | b_i=\beta),
    \label{eq:mapcri}
\end{equation}
for $i = 1, 2,\dotsc, k$, which is equivalent to
\begin{align}
    p(\boldsymbol{r} | b_i = 0) &\underset{\hat{b}_i = 0}{\overset{\hat{b}_i = 1}{\lessgtr}} p(\boldsymbol{r} | b_i = 1) \label{eq:compareMAP} \\
    \sum_{\bolds: b_i=0} p(\boldr|\bolds) &\underset{\hat{b}_i = 0}{\overset{\hat{b}_i = 1}{\lessgtr}} \sum_{\bolds: b_i=1} p(\boldr|\bolds). \label{eq:Bayesequiv}
\end{align}
The left-hand side of \eqref{eq:Bayesequiv} can be written as
\begin{equation}
    \sum_{\boldc: b_i=0} \prod_{j=1}^{n} \frac{1}{\sqrt{2\pi\sigma^2}} \exp\left( -\frac{(r_j-(2c_j-1))^2}{2\sigma^2} \right),
\end{equation}
which can be expanded, ignoring ${1}/{\sqrt{2\pi\sigma^2}}$, into
\begin{equation}\label{eq:beforeneglect} 
    \sum_{\boldc: b_i=0} 
    \exp\left(-\sum_{j=1}^{n} \frac{\cancel{r_j^2}-4r_jc_j +\cancel{2r_j}+\cancel{4c_j^2}-\cancel{4c_j}+\cancel{1}}{2\sigma^2}\right).
\end{equation}
Neglecting common terms (crossed out) in \eqref{eq:beforeneglect} yields
\begin{equation}
    \sum_{\boldc: b_i=0} \exp\left( \frac{2\boldsymbol{r}\boldsymbol{c}_\mathrm{T} }{\sigma^2}  \right) \underset{\hat{b}_i = 0}{\overset{\hat{b}_i = 1}{\lessgtr}} \sum_{\boldc: b_i=1} \exp\left( \frac{2 \boldsymbol{r}\boldsymbol{c}_\mathrm{T}}{\sigma^2}  \right), \label{eq:bitwisefinal}
\end{equation}
where $(\cdot)_\mathrm{T}$ indicates the transpose operation.
A decoder that performs \eqref{eq:bitwisefinal} for $i = 1, 2,\dotsc, k$ is said to be bit-wise optimal.

\subsection{Neural Networks}
In this work, the decoding of FEC codes via NNs is investigated.
An NN is a series of linear transformations of real-valued vectors, from a $1\times N_0$ input vector $\boldv_0$, to a $1\times N_{L+1}$ output vector $\boldv_{L+1}$. The transformations are defined via the recursive steps
\begin{equation}
    \boldv_i = \Gamma\left( \boldv_{i-1} \boldW_i + \boldb_i \right),
\end{equation}
for $i = 1, 2,\dotsc, L+1$, where $L$ is the number of hidden layers, and $N_i$ is the number of neurons in the $i^{\text{th}}$ layer.
The input and output layers correspond to $i=0$ and $i=L+1$, respectively.
Here, $\boldW_i$ is a $N_{i-1}\times N_{i}$ weight matrix, $\boldb_i$ is a $1\times N_i$ bias vector, and
\begin{equation}
    \Gamma(\boldv_i) = [\sigma(v_1), \sigma(v_2),\dotsc, \sigma(v_{N_i})],
\end{equation}
is a non-linear activation function.
The most common activation functions are the ReLU function 
\begin{equation}\label{eq.relu}
    \sigma(v_j) = \max(0,v_j),
\end{equation}
the sigmoid function 
\begin{equation}\label{eq.sigmoid}
    \sigma(v_j) = 1/(1+\exp(-v_j)),
\end{equation}
and the scaled softmax function
\begin{equation}\label{eq:scaledsoftmax}
     \sigma(v_j) =  \frac{\exp\left(\alpha v_j \right)}{\sum_{l=1}^{N_i} \exp(\alpha v_l)},
\end{equation}
where $\alpha$ is a scaling factor.

\section{Optimality of SLNN and MLNN Decoders}\label{sec:slnnmlnn}
In this section, we will derive optimal SLNN and MLNN architectures.
We will first show the results for the very simple Hamming $(7,4)$ code for illustrative purposes. At the end of this section, we will consider longer codes.

\subsection{Single-label Neural Network Decoders}\label{subsec:slnn}

\begin{figure}[t]
    \centering
     \includegraphics[width=\columnwidth]{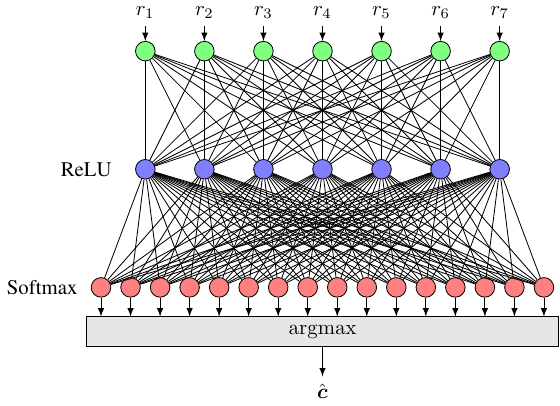}
    \caption{SLNN $7$-$7$-$16$ architecture used in~\cite{motani2023trans} to decode the Hamming~$(7,4)$ code and obtain near-optimal performance.}
    \label{fig:SLNNs}
\end{figure}

In~\cite{motani2019globecom, motani2023trans}, the decoding problem was formulated as a supervised single-label classification problem. 
The proposed SLNN decoder is a fully-connected NN with $N_0 = n$ input neurons, a hidden layer with $N_1$ neurons applying ReLU activation in \eqref{eq.relu}, and $N_{2} = 2^k$ output neurons using scaled softmax in \eqref{eq:scaledsoftmax} with $\alpha=1$ to output a probability distribution over $2^k$ possible codewords. 
The estimated transmitted codeword corresponds to the neuron that outputs the highest probability, determined by the $\argmax$ function, as shown in Fig.~\ref{fig:SLNNs}.

\begin{figure}[t]
\centering
\resizebox{\columnwidth}{!}{\begin{tikzpicture}
\begin{semilogyaxis}[
    xlabel={$\Eb/N_0$ (dB)},
    ylabel={FER},
    width=0.8\linewidth,  
    height=0.75\linewidth, 
    xmin=0,
    xmax=10,
    ymin=1e-6,
    ymax=1e-0,
    font=\footnotesize,
    mark options={solid},
    cycle list name=color list,
    enlargelimits=false,
    grid=major,
    grid style={dashed,lightgray!75},
    ylabel style={yshift=-0.5em},
    xlabel style={yshift=0.3em},
    xtick={0,2,...,10},
    ytick={1e-6,1e-5,1e-4,1e-3,1e-2,1e-1,1},
   legend style={font=\scriptsize,legend cell align=left, at={(0,1e-5)}, anchor=south west}, 
]

    \path (axis cs: 7,4.8e-2) node [anchor=south, black, rotate=-26] {Pre-FEC FER};
    
    \addplot[blue, dashed, thick, line width=1.2pt]
    table {data_SLNN_Hamming74_ML.dat}; 
    \label{sdml_fer_hamm_7_4}
    \legend{Maximum likelihood};

    \addplot[mark=*, mark size = 2pt, color=orange, no marks, dashdotted,line width=1.2pt,mark options={solid, purple, scale=.75, fill=white}] 
    table {data_SLNN7516_Hamming74.dat}; 
    \label{slnn_7_5_16_fer_hamm_7_4}
    \addlegendentry{SLNN $7$-$5$-$16$ \cite{motani2023trans}\phantom{x}}
    
    \addplot[mark=pentagon*, mark size = 3pt, color=red, line width=1.0pt, only marks, mark options={solid, red, scale=.75, fill=white}]
    table {data_SLNN7716_Hamming74.dat}; 
    \label{slnn_7_7_16_fer_hamm_7_4}
    \addlegendentry{SLNN $7$-$7$-$16$ \cite{motani2023trans}\phantom{x}}

    \addplot[mark=asterisk, mark size = 2.25pt, color=ForestGreen, line width=1.0pt, only marks, mark options={solid, ForestGreen, scale=.75, fill=white}]
    table {data_SLNN7016_Hamming74.dat}; 
    \label{slnn_7_16_fer_hamm_7_4}
    \addlegendentry{SLNN $7$-$0$-$16$ (Theorem~\ref{theorem1})}

    \addplot [mark=, color=black, solid, line width=1.2pt, forget plot]
    table[]{%
            0      0.66041
          0.5       0.6183
            1      0.57756
          1.5      0.53197
            2       0.4795
          2.5      0.43189
            3       0.3749
          3.5      0.32696
            4      0.27523
          4.5      0.22753
            5      0.18438
          5.5      0.14488
            6       0.1094
          6.5      0.08068
            7      0.05694
          7.5      0.03907
            8      0.02456
          8.5      0.01565
            9     0.008825
          9.5       0.0051
           10      0.00253
    }; 

\end{semilogyaxis}

\end{tikzpicture}} 
\caption{FER vs. $\Eb/N_0$ for different SLNN decoders for the Hamming~$(7,4)$ code. The results for SLNN $7$-$5$-$16$ and SLNN $7$-$7$-$16$ are based on our implementation of the architectures proposed in~\cite{motani2023trans}.}
\label{fig:SLNN_FERvsSNR}
\end{figure}
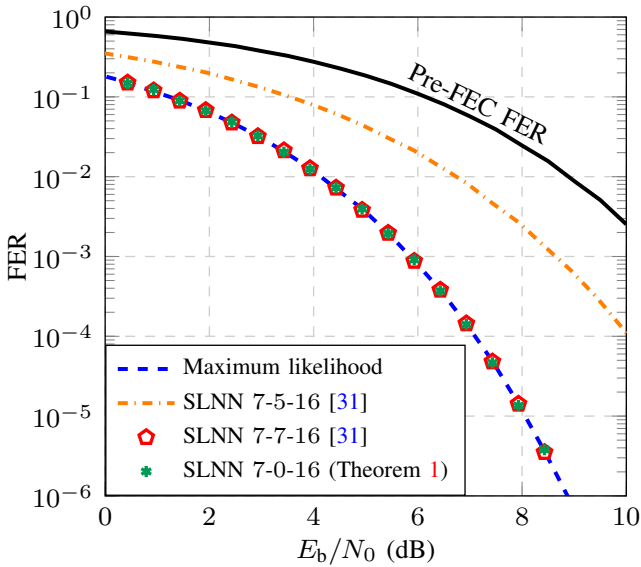

\begin{example}[SLNN for Hamming~$(7,4)$] 
The $7$-$7$-$16$ SLNN architecture used to decode the Hamming~$(7,4)$ code and achieve codeword-wise near-optimal decoding performance in~\cite[Fig. 2 (a)]{motani2023trans} is shown in Fig.~\ref{fig:SLNNs}.
The notation $N_0$-$N_1$-$N_2$ indicates the number of neurons in each layer.
The corresponding FER results are shown in Fig.~\ref{fig:SLNN_FERvsSNR}.\footnote{The SLNN decoder is trained based on a dataset that consists of pairs of one-hot encoded labels identifying the transmitted codewords and corresponding noisy received codewords. The dataset is generated via uniform random sampling from the codebook and transmission at $\Eb/N_0=0$~dB, which has been shown to generalize well across SNRs and achieve near-optimal performance~\cite{motani2023trans}.\label{ft:1}} 
The $7$-$7$-$16$ SLNN decoder achieves codeword-wise optimal performance.
On the other hand, the $7$-$5$-$16$ SLNN decoder performs significantly worse.
These results show that SLNN with a single hidden layer can approach optimal performance very closely, as long as the number of neurons in the hidden layer $N_1\geq n$. 
This observation agrees with~\cite[Fig. 2 (a)]{motani2023trans}.
\end{example}

\begin{figure}[t]
    \centering
    \includegraphics[width=\columnwidth]{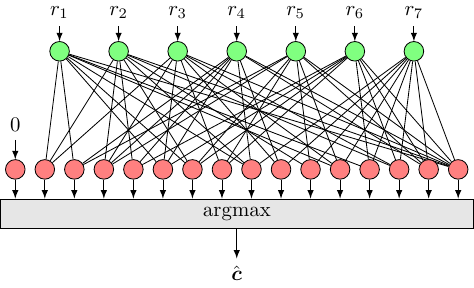}
    \caption{SLNN $7$-$0$-$16$ architecture proposed in Theorem~\ref{theorem1} to realize maximum likelihood decoding. The hidden layer in Fig.~\ref{fig:SLNN_FERvsSNR} is not present (it is not required for optimal performance). The number of edges here is $56$, while for SLNN $7$-$7$-$16$ in Fig.~\ref{fig:SLNN_FERvsSNR} the number of edges is  $161$.}
    \label{fig:SLNNthm}
\end{figure}

The following theorem is the first contribution of this paper.
It shows that the hidden layer in the SLNN architecture is, in fact, not needed. 
Optimal performance can always be guaranteed for any $(n,k)$ linear block code with a NN with $N_0=n$ input neurons and $N_2=2^k$ output neurons.

\begin{theorem}
    \label{theorem1}
    Let $(n,k)$ be a linear block code, with codebook $\mathcal{C}$ containing $2^k$ codewords. 
    Consider a two-layer NN with $n$ input neurons and $2^k$ output neurons. 
    Let the $n$-by-$2^k$ weight matrix $\boldsymbol{W}_1$ connecting the input layer to the output layer be binary, and has its columns equal to the codewords in $\mathcal{C}$. 
    This NN architecture is codeword-wise optimal. 
    No training is required.
\end{theorem}
\begin{IEEEproof}
The summation in \eqref{eq:mlcorr} can be expressed as a vector multiplication between $\boldsymbol{r}$ and $\boldsymbol{c}_\mathrm{T}^{(j)}$, i.e.,
\begin{equation}\label{eq:last-max}
    \hat{\boldsymbol{s}} 
    = \argmax_{\boldsymbol{s}^{(j)} \in \mathcal{S}} \boldsymbol{r} \boldsymbol{c}^{(j)}_\mathrm{T}.
\end{equation}
The proof is completed by representing this maximization problem in matrix form: Consider a vector-matrix multiplication between the received signal $\boldsymbol{r}$ and the $n$-by-$2^k$ matrix $\boldsymbol{W}_1$ for which each column is equal to a different $\boldsymbol{c} \in \mathcal{C}$. 
The $j^{\text{th}}$ element in the vector $\boldsymbol{r}\boldsymbol{W}_1$ is equal to the $j^{\text{th}}$ metric $\boldsymbol{r}\boldsymbol{c}^{(j)}_\mathrm{T}$ over which the maximization in~\eqref{eq:last-max} is carried out.
This is what the SLNN architecture with $N_0=n$, $N_1=0$, and $N_2=2^k$ realizes.
\end{IEEEproof}

Theorem~\ref{theorem1} shows that the optimal SLNN decoder, without any hidden layer, realizes codeword-wise maximum likelihood decoding. 
Furthermore, Theorem~\ref{theorem1} also implies that no training is required, as the (binary) weight matrix of the NN consists of the codewords in the codebook. 
The number of edges in the resulting NN is the sum of the Hamming weights of all codewords in $\mathcal{C}$.

\setcounter{example}{0}

\begin{example}[SLNN for Hamming~$(7,4)$ continued]\label{ex1}
Figure~\ref{fig:SLNNthm} shows the optimal SLNN $7$-$0$-$16$ architecture that realizes optimal decoding of the Hamming~$(7,4)$ code based on Theorem~\ref{theorem1}. 
We see that the neurons are sparsely connected.
The leftmost neuron in the output layer is, in fact, not connected at all, which is due to the fact that it corresponds to the all-zero codeword.
Figure~\ref{fig:SLNN_FERvsSNR} shows the corresponding FER performance. 
As Theorem~\ref{theorem1} predicted, the presented results show that the hidden layer in SLNN $7$-$7$-$16$ is indeed redundant: it only increases the computational complexity and memory requirements of the NN, and it also requires training. 
For this code, the number of edges in the NN is lowered from $161$ (for SLNN $7$-$7$-$16$) to only $56$ (for SLNN $7$-$0$-$16$).
\end{example}

\subsection{Multi-label Neural Network Decoders}\label{subsec:mlnn}

\begin{figure}[t]
    \centering
    \includegraphics[width=\columnwidth]{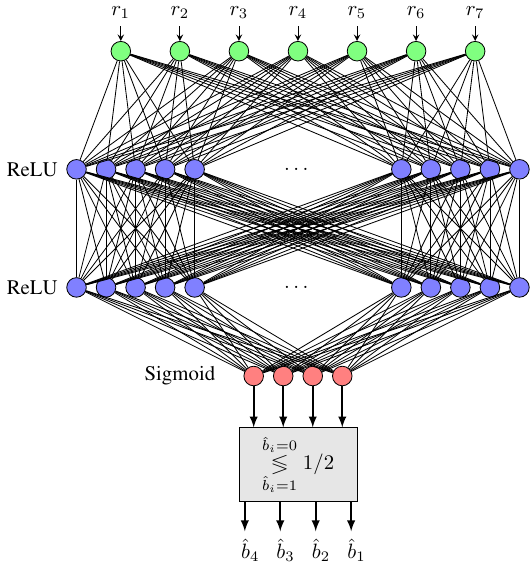}
    \caption{MLNN $7$-$50$-$50$-$4$ architecture used in~\cite{motani2023trans} to decode the Hamming~$(7,4)$ code and obtain near-optimal performance.}
    \label{fig:MLNNs}
\end{figure}

In \cite{motani2020icc, motani2020isit, motani2023trans}, the decoding problem was cast as a supervised multi-label classification problem. 
The proposed MLNN architecture is a fully-connected NN with $N_0 = n$ input neurons and $N_{L+1} = k$ output neurons applying sigmoid activation to output a probability distribution over each information bit.
The estimated information bits are determined by a thresholding operation as in Fig.~\ref{fig:MLNNs}.
The number of hidden layers $L$, the number of neurons $N_i$ for $i = 1, 2,\dotsc, L$ in the hidden layers, and the activation functions of the hidden layers are design parameters.

\begin{figure}[t]
\centering
\resizebox{\columnwidth}{!}{\begin{tikzpicture}
\begin{semilogyaxis}[
    xlabel={$\Eb/N_0$ (dB)},
    ylabel={BER},
    width=0.8\linewidth,  
    height=0.75\linewidth, 
    xmin=0,
    xmax=9,
    ymin=5e-8,
    ymax=1e-1,
    font=\footnotesize,
    mark options={solid},
    cycle list name=color list,
    enlargelimits=false,
    grid=major,
    xtick={0,2,...,10},
    grid style={dashed,lightgray!75},
    ylabel style={yshift=-0.5em},
    xlabel style={yshift=0.3em},
    ytick={1e-7,1e-6,1e-5,1e-4,1e-3,1e-2,1e-1},
      legend style={font=\scriptsize,legend cell align=left, at={(0,1e-5)}, anchor=south west}, 
]
    \path (axis cs: 7,7e-3) node [anchor=south, black, rotate=-22] {Pre-FEC BER};
    
    \addplot[mark=, color=black, solid, line width=1.2pt, forget plot] 
    table {data_MLNN_Hamming74_PreFEC.dat}; 

    \addplot[blue, dashed, thick, line width=1.2pt] 
    table {data_MLNN_Hamming74_ML.dat}; 
    \legend{Maximum likelihood};

    \addplot[mark=*, mark size = 2.75pt, color=red, line width=1.0pt, only marks, mark options={solid, red, scale=.75, fill=white}]
    table {data_MLNN750504_Hamming74.dat};
    \addlegendentry{MLNN $7$-$50$-$50$-$4$ \cite{motani2023trans}\phantom{x}}  

    \addplot[mark=triangle*, mark size = 2.75pt, color=ForestGreen, line width=0.7pt, only marks, mark options={solid, ForestGreen, scale=.75, fill=white}]
      table[]{%
                  0     0.080629
          0.5     0.065229
            1      0.05129
          1.5     0.039081
            2     0.028698
          2.5     0.020006
            3     0.013556
          3.5    0.0086633
            4    0.0052084
          4.5    0.0030452
            5    0.0015545
          5.5   0.00076104
            6   0.00034738
          6.5   0.00013746
            7   5.1604e-05
          7.5   1.6361e-05
            8   4.9873e-06
          8.5    1.174e-06
            9   2.1306e-07
      };
    \addlegendentry{MLNN $7$-$16$-$0$-$4$ (Theorem~\ref{theorem2})\phantom{x}}   

    \addplot [mark=asterisk, orange, only marks, mark size = 1.75pt, line width=0.5pt]
      table[]{%
       0	0.0815353585657371
    0.500000000000000	0.0657887146205156
    1	0.0516425803326336
    1.50000000000000	0.0393114391670232
    2	0.0287514032680554
    2.50000000000000	0.0200344493486920
    3	0.0135765736832446
    3.50000000000000	0.00866768994627782
    4	0.00520838595604951
    4.50000000000000	0.00304263171290405
    5	0.00155438084579103
    5.50000000000000	0.000763661876272127
    6	0.000347267066420664
    6.50000000000000	0.000138416099319183
    7	5.20143859608018e-05
    7.50000000000000	1.63809988279622e-05
    8	5.10123999372155e-06
    8.50000000000000	1.16625148113938e-06
    9	2.13055303717135e-07          
    }; 
    \addlegendentry{MLNN $7$-$16$-$0$-$4$ (Theorem~\ref{theorem2}, fixed $\alpha$)}

\end{semilogyaxis}

\end{tikzpicture}}
\caption{BER vs. $\Eb/N_0$ for different MLNN decoders for the Hamming~$(7,4)$ code. The results for MLNN $7$-$50$-$50$-$4$ are based on our implementation of the architectures proposed in~\cite{motani2023trans}.}
\label{fig:MLNN_BERvsSNR}
\end{figure}
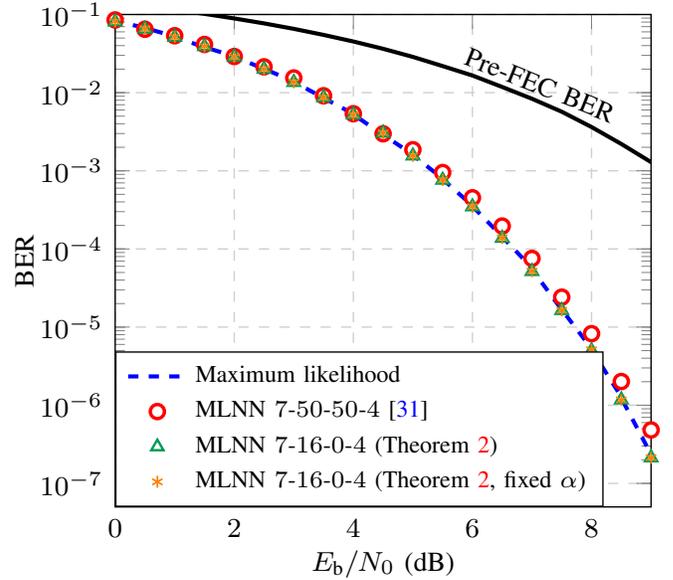

\begin{example}[MLNN for Hamming~$(7,4)$]
The $7$-$50$-$50$-$4$ MLNN architecture used to decode the Hamming $(7,4)$ code and achieve near-optimal bit-wise decoding performance in~\cite[Fig. 5 (a)]{motani2023trans} is shown in Fig.~\ref{fig:MLNNs}.
This network uses ReLU activation in \eqref{eq.relu} in the hidden layers~\cite[Table III]{motani2023trans}.
The corresponding BER results are shown in Fig.~\ref{fig:MLNN_BERvsSNR}.\footnote{The MLNN decoder is trained based on a dataset that consists of pairs of information bit vectors that are encoded into the transmitted codewords and the corresponding noisy received codewords.} 
The $7$-$50$-$50$-$4$ MLNN decoder approaches bit-wise optimal performance very closely.
On the other hand, the MLNN $7$-$100$-$4$ decoder, i.e., MLNN with a single hidden layer, performs significantly worse, as previously shown in~\cite[Fig. 5 (a)]{motani2023trans}.
\end{example}

The following theorem is the second contribution of this paper.
It shows that there exists an MLNN architecture with a single hidden layer that realizes bit-wise maximum likelihood decoding for any $(n,k)$ linear block code with $N_0=n$, $N_1=2^k$, and $N_2=k$ input, hidden, and output neurons, respectively.

\begin{theorem}
    \label{theorem2}
    Let $(n,k)$ be a linear block code $\mathcal{C}$. 
    Consider a three-layer NN with $n$ input neurons, $2^k$ hidden neurons applying the scaled softmax in \eqref{eq:scaledsoftmax} with $\alpha=2/\sigma^2$ as activation function, and $k$ output neurons. 
    Let the $n$-by-$2^k$ weight matrix $\boldsymbol{W}_1$ connecting the input layer to the hidden layer be binary, and has its columns equal to the codewords in $\mathcal{C}$.
    Let the $2^k$-by-$k$ weight matrix $\boldsymbol{W}_2$ connecting the hidden layer to the output layer be binary, and has its rows equal to all possible $k$-bit input strings of the code.
    We assume that the columns and rows of $\boldsymbol{W}_1$ and $\boldsymbol{W}_2$, respectively, are sorted such that $j^{\text{th}}$ row of $\boldsymbol{W}_2$ is encoded into $j^{\text{th}}$ column of $\boldsymbol{W}_1$ for $j = 1, 2,\dotsc, 2^k$.
    This NN architecture is bit-wise optimal.
    No training is required.
\end{theorem}

\begin{IEEEproof}
Dividing both sides of \eqref{eq:bitwisefinal} by the sum of both sides, we get
\begin{equation}
    \sum_{\substack{\boldsymbol{c}: b_i = 0}}  \frac{\exp\left( \frac{2 \boldsymbol{r}\boldsymbol{c}_\mathrm{T}}{\sigma^2} \right)}{\sum_{\boldc} \exp\left( \frac{2\boldsymbol{r}\boldsymbol{c}_\mathrm{T}}{\sigma^2} \right)} \underset{\hat{c}_i = 0}{\overset{\hat{c}_i = 1}{\lessgtr}}  \sum_{\substack{\boldsymbol{c}:b_i = 1}} \frac{\exp\left( \frac{2\boldsymbol{r}\boldsymbol{c}_\mathrm{T}}{\sigma^2} \right)}{\sum_{\boldc} \exp\left( \frac{2\boldsymbol{r}\boldsymbol{c}_\mathrm{T}}{\sigma^2} \right)}.
    \label{eq:comparisontheorem2}    
\end{equation}
In the NN defined in Theorem~\ref{theorem2}, the input-to-hidden transformation with the scaled softmax activation \eqref{eq:scaledsoftmax} computes the addends of the outer summations on both sides of \eqref{eq:comparisontheorem2}.
Then the hidden-to-output transformation computes the outer summation on the right-hand side.
Observing also that the left and right sides of \eqref{eq:comparisontheorem2} add up to $1$, a thresholding operation applied to the output nodes of the NN, with threshold set at $1/2$, results in \eqref{eq:bitwisefinal}, and thus in \eqref{eq:mapcri}. 
\end{IEEEproof}

Theorem~\ref{theorem2} shows that the optimal MLNN decoder, with a single hidden layer, realizes bit-wise maximum likelihood decoding. 
This is not surprising since the universal approximation theorem implies that feedforward NNs having a non-polynomial activation function with as few as one hidden layer are universal approximators.
Furthermore, Theorem~\ref{theorem2} also indicates that no training is required, as the (binary) weight matrices of the NN consist of the possible inputs and codewords of the code. 
The number of edges that connect the input layer to the hidden layer in the resulting NN is the sum of the Hamming weights of all codewords in $\mathcal{C}$.
The number of edges that connect the hidden layer to the output layer in the resulting NN is the sum of the Hamming weights of all possible input strings, which is $k2^{k-1}$.

\setcounter{example}{1}

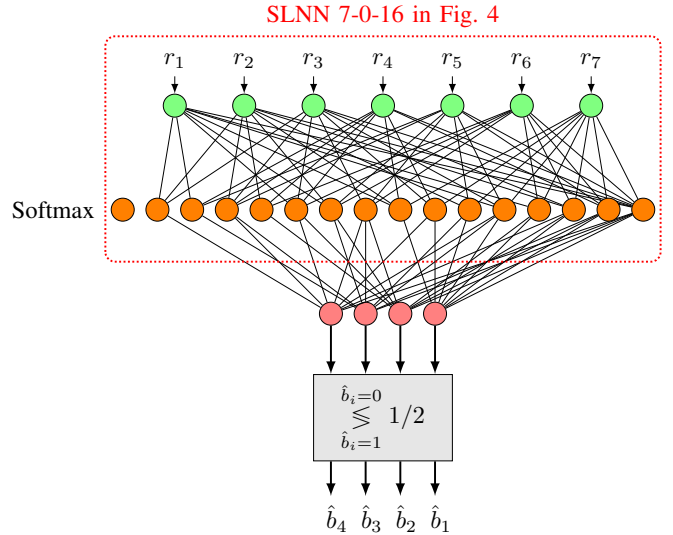
\begin{figure}[t]
    \centering
    \resizebox{\columnwidth}{!}{ \begin{tikzpicture}[
        input/.style={circle, draw=black},
        output/.style={circle, draw=black},
        label/.style={above=0.25cm},    
        block/.style={rectangle, draw, minimum width=2cm, minimum height=0.5cm, fill=gray!20}  
    ]
    \foreach \i/\text in {1/$r_1$, 2/$r_2$, 3/$r_3$, 4/$r_4$, 5/$r_5$, 6/$r_6$, 7/$r_7$} {
        \node[input, fill=green!50] (I-\i) at (1*\i+0.25,0) {};
        \node[label] (label-\i) at (I-\i.north) {\scalebox{1}{\text}};  
        \draw[-latex] (label-\i.south) -- (I-\i.north);  
    }
    
    \foreach \o [count=\oi] in {1, 2, 3, 4, 5, 6, 7, 8, 9 ,10, 11, 12, 13, 14, 15, 16} {
        \node[output, draw=none, fill=white] (O-\oi) at (0.5*\oi,-1.5) {};
    }

    \foreach \o [count=\oi] in {1, 2, 3, 4} {
        \node[output, fill=red!50] (H-\oi) at (3+0.5*\oi,-3) {};
    }
    
    \draw[-] (I-1) -- (O-2);
    \draw[-] (I-2) -- (O-2);
    \draw[-] (I-3) -- (O-2);

    \draw[-] (I-1) -- (O-3);
    \draw[-] (I-4) -- (O-3);
    \draw[-] (I-5) -- (O-3);

    \draw[-] (I-2) -- (O-4);
    \draw[-] (I-3) -- (O-4);
    \draw[-] (I-4) -- (O-4);
    \draw[-] (I-5) -- (O-4);

    \draw[-] (I-2) -- (O-5);
    \draw[-] (I-4) -- (O-5);
    \draw[-] (I-6) -- (O-5);

    \draw[-] (I-1) -- (O-6);
    \draw[-] (I-3) -- (O-6);
    \draw[-] (I-4) -- (O-6);
    \draw[-] (I-6) -- (O-6);

    \draw[-] (I-1) -- (O-7);
    \draw[-] (I-2) -- (O-7);
    \draw[-] (I-5) -- (O-7);
    \draw[-] (I-6) -- (O-7);

    \draw[-] (I-3) -- (O-8);
    \draw[-] (I-5) -- (O-8);
    \draw[-] (I-6) -- (O-8);

    \draw[-] (I-1) -- (O-9);
    \draw[-] (I-2) -- (O-9);
    \draw[-] (I-4) -- (O-9);
    \draw[-] (I-7) -- (O-9);

    \draw[-] (I-3) -- (O-10);
    \draw[-] (I-4) -- (O-10);
    \draw[-] (I-7) -- (O-10);

    \draw[-] (I-2) -- (O-11);
    \draw[-] (I-5) -- (O-11);
    \draw[-] (I-7) -- (O-11);

    \draw[-] (I-1) -- (O-12);
    \draw[-] (I-3) -- (O-12);
    \draw[-] (I-5) -- (O-12);
    \draw[-] (I-7) -- (O-12);

    \draw[-] (I-1) -- (O-13);
    \draw[-] (I-6) -- (O-13);
    \draw[-] (I-7) -- (O-13);

    \draw[-] (I-2) -- (O-14);
    \draw[-] (I-3) -- (O-14);
    \draw[-] (I-6) -- (O-14);
    \draw[-] (I-7) -- (O-14);

    \draw[-] (I-4) -- (O-15);
    \draw[-] (I-5) -- (O-15);
    \draw[-] (I-6) -- (O-15);
    \draw[-] (I-7) -- (O-15);

    \draw[-] (I-1) -- (O-16);
    \draw[-] (I-2) -- (O-16);
    \draw[-] (I-3) -- (O-16);
    \draw[-] (I-4) -- (O-16);
    \draw[-] (I-5) -- (O-16);
    \draw[-] (I-6) -- (O-16);
    \draw[-] (I-7) -- (O-16);

    \draw[-] (O-2) -- (H-1);
    \draw[-] (O-3) -- (H-2);
    \draw[-] (O-4) -- (H-1);
    \draw[-] (O-4) -- (H-2);
    
    \draw[-] (O-5) -- (H-3);
    \draw[-] (O-6) -- (H-1);
    \draw[-] (O-6) -- (H-3);
    \draw[-] (O-7) -- (H-2);
    \draw[-] (O-7) -- (H-3);
    \draw[-] (O-8) -- (H-1);
    \draw[-] (O-8) -- (H-2);
    \draw[-] (O-8) -- (H-3);
    
    \draw[-] (O-9) -- (H-4);
    \draw[-] (O-10) -- (H-1);
    \draw[-] (O-10) -- (H-4);
    \draw[-] (O-11) -- (H-2);
    \draw[-] (O-11) -- (H-4);
    \draw[-] (O-12) -- (H-1);
    \draw[-] (O-12) -- (H-2);
    \draw[-] (O-12) -- (H-4);
    
    \draw[-] (O-13) -- (H-3);
    \draw[-] (O-13) -- (H-4);
    \draw[-] (O-14) -- (H-1);
    \draw[-] (O-14) -- (H-3);
    \draw[-] (O-14) -- (H-4);
    \draw[-] (O-15) -- (H-2);
    \draw[-] (O-15) -- (H-3);
    \draw[-] (O-15) -- (H-4);
    \draw[-] (O-16) -- (H-1);
    \draw[-] (O-16) -- (H-2);
    \draw[-] (O-16) -- (H-3);
    \draw[-] (O-16) -- (H-4);

    \node[block] (argmax) at (4.25,-4.5) {\scalebox{1}{$\underset{\hat{b}_i = 1}{\overset{\hat{b}_i = 0}{\lessgtr}} 1/2$}};

    \foreach \i in {1, 2, 3, 4} {
        \draw[latex-,thick] (argmax.north) ++(-1.25+\i*0.5, 0) -- (H-\i);
    } 
    
    \foreach \i in {1, 2, 3, 4} {
        \draw[-latex,thick] (argmax.south) ++(1.25-\i*0.5, 0) -- ++(0,-0.5) node[below] {\scalebox{1}{\hphantom{x}$\hat{b}_{\i}$}}; 
    } 

     \draw[densely dotted, red, thick, rounded corners] (0.25,-2.25) rectangle (8.25,1);
     \path (0.25,1) --  (8.25,1) node[midway,above,red] {SLNN 7-0-16 in Fig.~\ref{fig:SLNNthm}};
     
    \foreach \o [count=\oi] in {1, 2, 3, 4, 5, 6, 7, 8, 9 ,10, 11, 12, 13, 14, 15, 16} {
        \node[output, fill=orange] (O-\oi) at (0.5*\oi,-1.5) {};
    }

    \node at (-0.5, -1.5) {\scalebox{1}{Softmax}};
     
\end{tikzpicture}}
    \caption{MLNN $7$-$16$-$4$ architecture proposed in Theorem~\ref{theorem2} to realize maximum likelihood decoding. The number of edges here is $88$, unlike in Fig.~\ref{fig:MLNNs}, where $3200$ edges are required.}
    \label{fig:MLNNthm}
\end{figure}

\begin{example}[MLNN for Hamming~$(7,4)$ continued]
Figure~\ref{fig:MLNNthm} shows the optimal MLNN $7$-$16$-$4$ architecture that realizes optimal decoding of the Hamming $(7,4)$ code based on Theorem~\ref{theorem2}.
We see that the neurons are sparsely connected.
The leftmost neuron in the output layer is not connected at all, similar to Example~\ref{ex1}.
Figure~\ref{fig:MLNN_BERvsSNR} shows the corresponding BER performance.
As Theorem~\ref{theorem2} predicted, the presented results show that the second hidden layer and most neurons in the first hidden layer in MLNN $7$-$50$-$50$-$4$ are redundant.
For this code, the number of edges in the NN is reduced from $3200$ (for MLNN $7$-$50$-$50$-$4$) to only $88$ (for MLNN $7$-$16$-$4$).
We note that the $100$ hidden nodes of MLNN $7$-$50$-$50$-$4$ in~\cite{motani2023trans} apply ReLU activation.
On the other hand, the $16$ hidden nodes in our MLNN $7$-$16$-$4$ apply softmax, which is computationally more complex, with an SNR-dependent scaling factor $\alpha=2/\sigma^2$.
However, it can be seen from Fig.~\ref{fig:MLNN_BERvsSNR} that performance is not sensitive to mismatch in $\alpha$.
When $\alpha$ computed for $\Eb/N_0=4$~dB used for all SNRs, BERs are virtually the same.
\end{example}

We note that the NN described in Theorem~\ref{theorem1} corresponds to the first two layers of the NN described in Theorem~\ref{theorem2} as shown within a dotted red rectangle in Fig.~\ref{fig:MLNNthm}.
Thus, the MLNN proposed in Theorem~\ref{theorem2} can be seen as a ``concatenation'' of the SLNN proposed in Theorem~\ref{theorem1} and an output layer.
Finally, we emphasize that we are not using Theorems~\ref{theorem1} and~\ref{theorem2} to propose that the SLNN and MLNN decoders are the solution to the FEC decoding problem.
We are merely demonstrating that what was proposed in the literature based on heuristic design rules is not optimal in terms of complexity or performance. 

\subsection{Scalability to Longer Codes}\label{sec:scale}

\begin{figure*}[t]
\centering
\addtocounter{subfigure}{-1} 
\subfloat{\centering
             \resizebox{0.9\textwidth}{!}{\begin{tikzpicture}
            	\begin{axis}[
            		width=1\textwidth,
            		hide axis,
            		xmin=0,
                    xmax=0,
                    ymin=0,
                    ymax=0,
                    legend style={font=\tiny,row sep=-0.25ex,inner sep=0.2ex},
                    legend columns = 6,
                     legend cell align={left}
            		]

                    \addlegendimage{black, thick, line width=1.2pt}
                    \addlegendentry{Pre-FEC\phantom{xx}}
            
                    \addlegendimage{blue, dashed, thick, line width=1.2pt}
                    \addlegendentry{Maximum likelihood\phantom{xx}}
            
                    \addlegendimage{mark=triangle*, red, mark size = 2.25pt, line width=1pt, mark options={solid, red, fill=white}}
                    \addlegendentry{BDD\phantom{xx}}
            
                    \addlegendimage{mark=diamond, orange, only marks, mark size = 2.25pt, line width=1pt}\label{thm1mark}
                    \addlegendentry{Theorem~\ref{theorem1}\phantom{xx}}
            
                    \addlegendimage{mark=square*, only marks, mark size = 2pt, line width=1pt, mark options={solid, orange, fill=white}}\label{thm2mark}
                    \addlegendentry{Theorem~\ref{theorem2}\phantom{xx}}           
            
            		\addlegendimage{mark=asterisk, dashed, ForestGreen, mark size = 2pt, line width=1pt, mark options={solid, ForestGreen, fill=white}}\label{greenmark}
                    \addlegendentry{MLNN $31$-$2000$-$1000$-$21$~\cite[Fig. 8 (d)]{motani2023trans}}

            	\end{axis}  
            \end{tikzpicture}}
}\\
\vspace{-0.4cm}
\subfloat[FER for polar $(16,8)$.]{\label{fig:ferpolar}\includegraphics[width=0.25\textwidth]{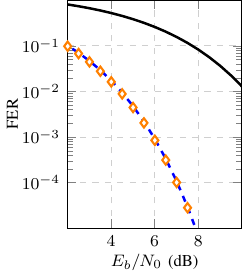}}
\subfloat[BER for polar $(16,8)$.]{\label{fig:berpolar}\includegraphics[width=0.25\textwidth]{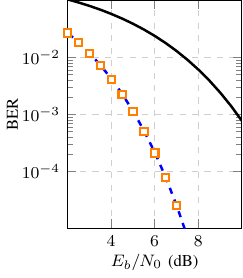}}
\subfloat[FER for BCH $(31,21)$.]{\label{fig:ferbch}\includegraphics[width=0.25\textwidth]{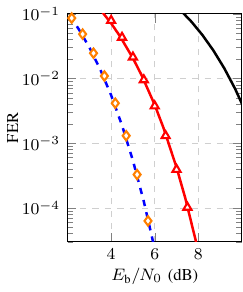}}
\subfloat[BER for BCH $(31,21)$.]{\label{fig:berbch}\includegraphics[width=0.25\textwidth]{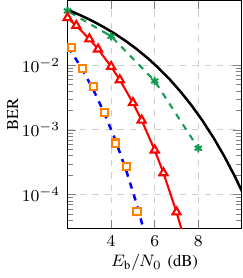}}
\caption{FER/BER vs. $\Eb/N_0$ of Polar $(16,8)$ and BCH $(31,21)$ codes for SLNN/MLNN and BDD decoders.}
\label{fig:waterfallspolarbch}
\end{figure*}

As we discussed in Sec.~\ref{sec:intro}, scalability of NN decoders for longer FEC codes is hindered by the curse of dimensionality, i.e., the size of networks that provide adequate performance increases exponentially with $k$.
This is indeed the case for the SLNN and MLNN decoders as they both have at least one layer where the number of neurons grows exponentially with $k$.
Here, we study the performance of NNs defined in Theorems~\ref{theorem1} and~\ref{theorem2} for the $(16,8)$ polar code considered in~\cite{dlbasedchanneldecoding} and for the $(31,21)$ BCH code considered in~\cite{motani2023trans}. 
The properties of the corresponding NNs are given in Table~\ref{tab:NNs}.
We see from Fig.~\ref{fig:waterfallspolarbch} again that the NNs defined in Theorems~\ref{theorem1} and~\ref{theorem2} realize codeword- and bit-wise maximum likelihood decoding, respectively, for both codes.

In the case of the $(16,8)$ polar code, the MLNN used in~\cite{dlbasedchanneldecoding} was also able to obtain bit-wise optimal performance. 
We also note that the MLNN used in~\cite[Fig. 4]{motani2021nips} was unable to achieve optimal performance for a $(16,8)$ polar code, although the objective there was to reduce complexity by pruning rather than to obtain the best performance.
We see from Table~\ref{tab:NNs} that both these MLNNs have a larger size than the optimal NNs described in Theorems~\ref{theorem1} and~\ref{theorem2}, and require training.

The results for the $(31,21)$ BCH code are shown in Fig.~\ref{fig:ferbch} and Fig.~\ref{fig:berbch}. In these figures, we also show the performance obtained by a very simple traditional hard-decision decoder: BDD (see Fig.~\ref{fig:commsyst}). The performance of the MLNN decoder in~\cite[Fig. 8 (d)]{motani2023trans} is shown in Fig.~\ref{fig:berbch} (green stars) and is worse than BDD. These results show that heuristically designed MLNN decoders are not competitive when compared to traditional decoders.\footnote{This conclusion also holds for other NN-based decoders such as the error correction code transformer~\cite{ECCT} and the cross-attention message
passing transformer~\cite{park2024crossmptcrossattentionmessagepassingtransformer} as shown in \cite[Sec. IV]{scheepers2025mlbaseddec}.} 
As expected, the results from Theorems~\ref{theorem1} and \ref{theorem2} match those of maximum likelihood. However, as shown in the second to last column of Table~\ref{tab:NNs}, this comes at the cost of significant complexity.

\setlength{\tabcolsep}{3pt}
\begin{table}[t]
\caption{Properties of NN Decoders Considered in Sec.~\ref{sec:scale} and in Fig.~\ref{fig:waterfallspolarbch}}
\label{tab:NNs}
\renewcommand{\arraystretch}{1.3}
\centering
\resizebox{\columnwidth}{!}{
\begin{tabular}{|c|ccccc|} 
\hline
                       & Ref.             & NN                       & Train & \# Edges & Performance        \\
\hline\hline
\multirow{4}{*}{\rotatebox[origin=c]{90}{Polar}} & \cite{dlbasedchanneldecoding} & $2^4$-$2^7$-$2^6$-$2^5$-$2^3$ & Yes      & $12544$, real & Bit-wise optimal  \\
 & \cite{motani2021nips}         & $2^4$-$2^8$-$2^8$-$2^4$    & Yes      & $<71689^{\dagger}$, real & Suboptimal \\
 & \ref{thm1mark} & $2^4$-$2^8$       & No & $2^{11}$, binary & Codeword-wise optimal \\
 & \ref{thm2mark} & $2^4$-$2^8$-$2^3$ & No & $3072$, binary & Bit-wise optimal  \\
\hline
 \multirow{3}{*}{\rotatebox[origin=c]{90}{BCH}} & \ref{greenmark}         & $31$-$2000$-$1000$-$21$  & Yes      & $2.08$M, real & Worse than BDD  \\
  & \ref{thm1mark} & $31$-$2^{21}$      & No & $32.5$M, binary & Codeword-wise optimal \\
  & \ref{thm2mark} & $31$-$2^{21}$-$21$ & No & $54.5$M, binary & Bit-wise optimal \\
\hline  
\multicolumn{6}{c}{\small $^{\dagger}$\cite{motani2021nips} starts from a $2^4$-$2^8$-$2^8$-$2^4$ NN with $71689$ edges, and achieves }\\ 
\multicolumn{6}{c}{\small complexity reduction via weight pruning at the cost of a loss in performance.} \\
\end{tabular}
}
\end{table}

\section{Conclusions}\label{sec:conclusion}
We provided a formal analysis of two neural network decoders for error-correcting codes: SLNN and MLNN decoders. 
These decoders have been shown in the literature to achieve codeword- and bit-wise {\it near}-optimal decoding performance, respectively. The designs in the literature are based on heuristic rules. 
In this paper, we showed analytically that these decoders, in fact, always realize optimal decoding for certain network architectures, without even requiring any training. 
This optimality was also numerically demonstrated for the Hamming $(7,4)$, polar $(16,8)$, and BCH $(31,21)$ codes.
Moreover, these optimal networks have binary weight matrices and are more sparsely connected than the ones achieving near-optimal performance.

Our optimal networks require at least one layer with size $2^k$. 
Thus, our overall conclusion is that it is possible to achieve optimal decoding for codes with $k<32$ with a ``trivial'' NN that does not require training.
Moreover, the complexity of these trivial NNs is smaller than that of ``actual'' suboptimal NNs that require training of real-valued weight matrices and bias vectors.
On the other hand, for longer codes ($k>32$), NNs that can achieve optimal performance quickly grow to impractical sizes, while smaller networks provide significantly poor performance.

\bibliographystyle{IEEEtran}
\bibliography{tgcn-references}

\end{document}